\documentclass[journal]{IEEEtran}
\usepackage{cite}

\usepackage{subfig}
\usepackage{graphicx}
\usepackage{lineno}

\ifCLASSINFOpdf
\else
\fi

\begin{document}
%
\title{The ATLAS Tile Calorimeter performance and its upgrade towards the High-Luminosity LHC}
%
%
%

\author{Merve Nazlim Agaras,~\IEEEmembership{on behalf of the ATLAS Collaboration}

\thanks{
Copyright 2021 CERN for the benefit of the ATLAS Collaboration. Reproduction of this article or parts of it is allowed as specified in the CC-BY-4.0 license.
}
\thanks{M.Nazlim Agaras is with Institut de Fisica d’Altes Energies (IFAE),
Campus UAB, Facultat Ciencies Nord, 08193 Bellaterra, Barcelona, Spain. (e-mail:merve.nazlim.agaras@cern.ch)}
}

\maketitle

\begin{abstract}
The Tile Calorimeter (TileCal) is a sampling hadronic calorimeter covering the central region of the ATLAS experiment. TileCal uses steel as absorber and plastic scintillators as active medium. The scintillators are read-out by the wavelength shifting fibres coupled to the photomultiplier tubes (PMTs). The analogue signals from the PMTs are amplified, shaped, digitized by sampling the signal every 25 ns and stored on detector until a trigger decision is received. The TileCal front-end electronics reads out the signals produced by about 10000 channels measuring energies ranging from about 30 MeV to about 2 TeV. Each stage of the signal production from scintillation light to the signal reconstruction is monitored and calibrated to better than 1\% using radioactive source, laser and charge injection systems. The performance of the calorimeter has been measured and monitored using calibration data, cosmic ray muons and the large sample of proton-proton collisions acquired in 2009-2018 during LHC Run I and Run II.
The High-Luminosity phase of LHC, delivering five times the LHC nominal instantaneous luminosity, is expected to begin in 2028. TileCal will require new electronics to meet the requirements of a higher trigger rate, higher ambient radiation, and to ensure better performance under high pile-up conditions. Both the on- and off-detector TileCal electronics will be replaced during the shutdown of 2025-2027. PMT signals from every TileCal cell will be digitized and sent directly to the back-end electronics, where the signals are reconstructed, stored, and sent to the first level of trigger at a rate of 40 MHz. This will provide better precision of the calorimeter signals used by the trigger system and will allow the development of more complex trigger algorithms. Changes to the electronics will also contribute to the data integrity and reliability of the system. New electronics prototypes were tested in laboratories as well as in beam tests.
Results of the calorimeter calibration and performance during LHC Run II are summarized, the main features and beam test results obtained with the new front-end electronics are also presented.
\end{abstract}


%
\IEEEpeerreviewmaketitle

\section{Introduction}
\begin{figure}[h!]
\centering
\subfloat[\label{fig:detector}]{\includegraphics[width=0.50\textwidth]{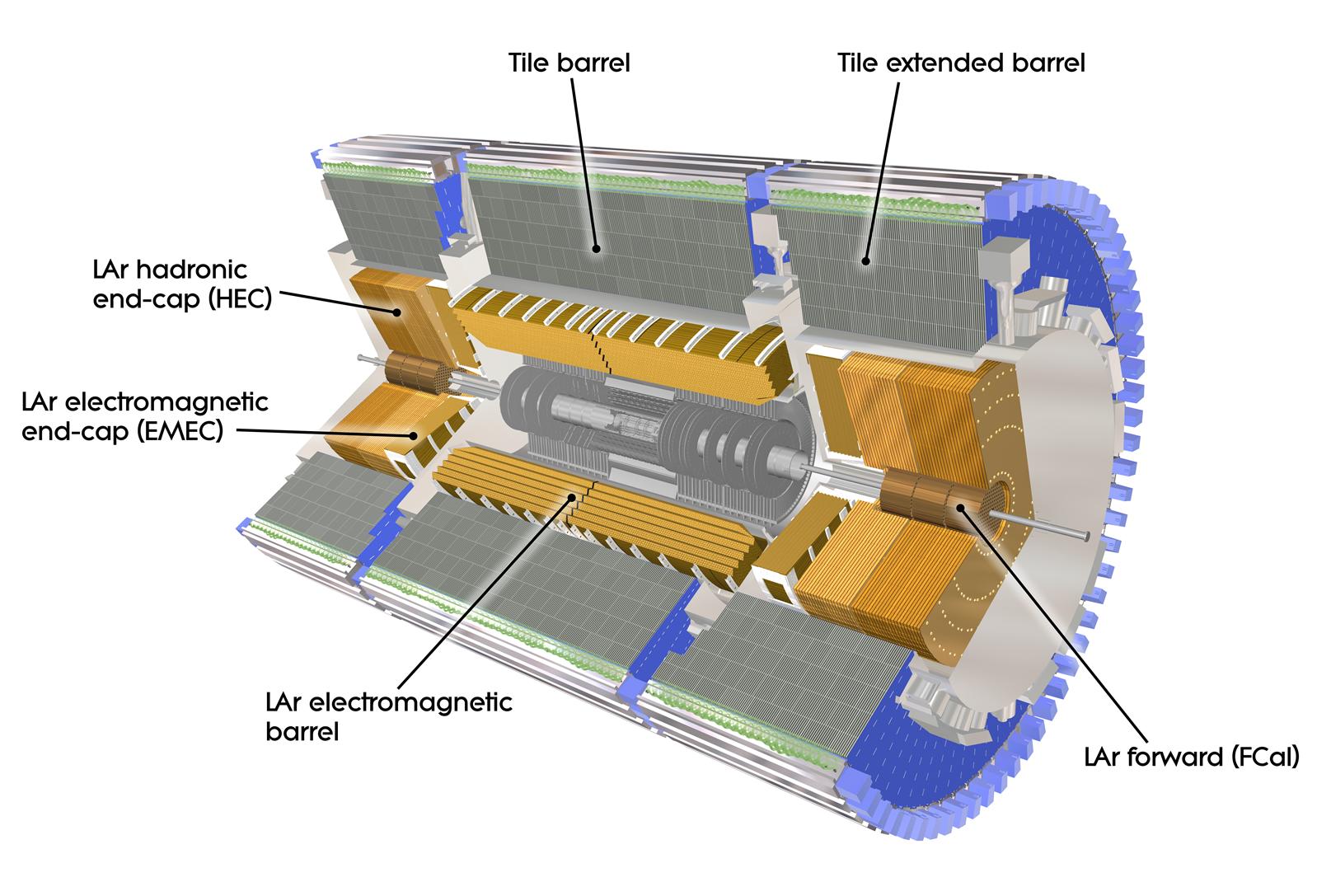}}\\
  \caption{ Schematic view of ATLAS detector system~\cite{public}.\label{fig:tile}}
\end{figure}

%
%
%
%
The ATLAS detector \cite{PERF-2007-01} surrounds the interaction region at Point 1 of the Large Hadron Collider (LHC) and is one of the two general purpose particle detectors (Figure~\ref{fig:tile}). It is 25~m in height and 44~m in length, with an overall weight of approximately 7000 tonnes. The purpose of ATLAS is to identify the particles, and measure the properties of elementary particles produced in proton-proton collisions (pp) or heavy ions collisions.
To do so, the detector consists of multiple sub-detectors with good hermiticity. The hadronic Tile calorimeter (TileCal)~\cite{TCAL-2010-01} is one the ATLAS sub-detectors covering the barrel region (pseudo-rapidity $|\eta|<$ 1.7). It contributes to the energy measurements of particles and jets produced in a center of mass energy up to  $\sqrt{s}=13$~TeV proton-proton interactions as well as to the missing transverse energy measurements, jet substructure, electron isolation and triggering.

TileCal is a sampling calorimeter composed of scintillating tiles as active material and steel plates as absorber. It is divided into three cylinders: Long-Barrel (LB) located in the ATLAS barrel, covers the region $|\eta| < 1.0$, while the others are in the endcaps, called Extended-Barrel (EB), covers the region $0.8 < |\eta| < 1.7$. Tile cells and electronics are organised into 4 partitions, LBA and LBC for the A-side and C-side of the barrel region, and separate EBA and EBC partitions in the extended barrel region. 
The scintillation light is collected at the edges of each tile from the two opposite sides by wavelength-shifting (WLS) fibres, arranged in groups defining the readout cells, and connected to a pair of photomultiplier tubes (PMTs) within a module to increase the uniformity of the response and the reliability of the light collection. There are 45 cells in each LB module and 32 cells in each EB module, read by around 10,000 PMTs (or named as \emph{channel}) in total. 

\section{Signal reconstruction and calibration systems}



Light collected by each PMT makes a current pulse whose amplitude is proportional to the energy deposited by traversing particles in the associated cell. The output pulse is shaped and amplified by the electronics located on the detector close to the PMTs. Later the shaped signal is amplified into two analogue pulses: high and low gain, which are then digitized by two 10-bit ADCs. The resulting pulse shape is then represented by seven samples, which  are waiting for first level trigger to be sent to the back-end electronics.

\begin{figure}[h!]
\centering
\includegraphics[width=0.5\textwidth]{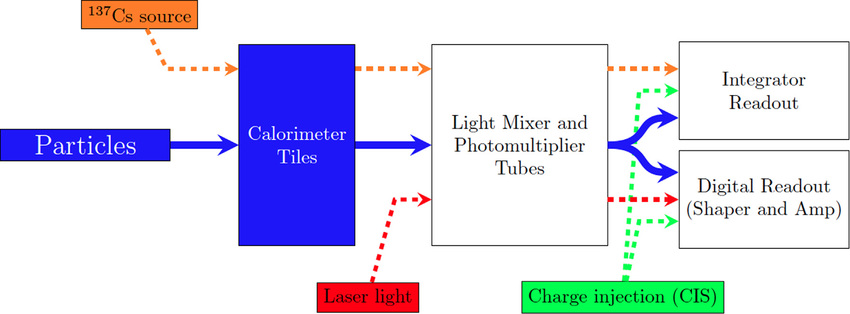}
  \caption{ Schematic representation of the readout signal path from different TileCal calibration systems.\label{fig:calib}}
\end{figure}

Several calibration systems are used to monitor the stability of the different stages of the signal processing elements of TileCal. These systems provide per channel calibration constants.
Figure~\ref{fig:calib} shows the different calibration systems along with the signal path.
The paths of different monitoring and calibration systems are relatively overlapping with each other, implying that some parts of the detector can be monitored by more than one system. This allows for validation of potential problems from different monitoring systems.

\begin{figure}[h!]
\centering
\includegraphics[width=0.4\textwidth]{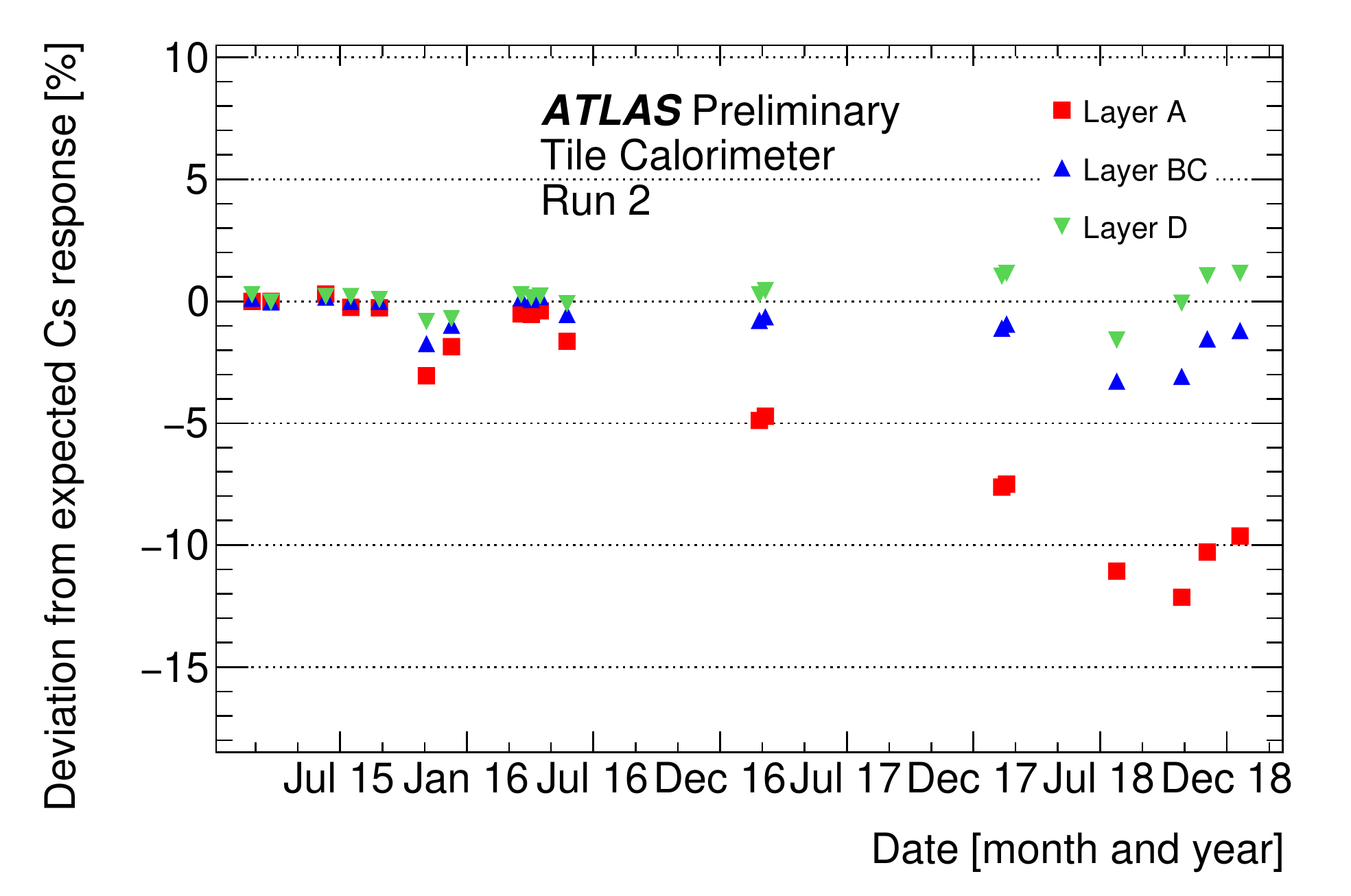}
  \caption{The variation in the TileCal response measured by the Cs system during Run II for different layers~\cite{public}.\label{fig:ces}}
\end{figure}

\hspace{1pt}
\paragraph{Cesium calibration system} \hspace{0pt} \\
The TileCal uses three radioactive $^{137}Cs$ sources, which are displaced using a hydraulic system to scan all the individual scintillator tiles in order to sustain the global electromagnetic scale response and to monitor all optics components and PMTs. Any deviation of TileCal response to Cesium signals from the reference signal can be interpreted as a scintillator degradation or PMT gain variations and translated into calibration constants. The variation in the TileCal response measured by the Cs system during Run II is given in Figure~\ref{fig:ces} and the biggest drift is observed in the layer A, which is the closest to the collision point. The precision of the system is at the order of 0.3\% for one typical cell. Between Run I and Run II, the stability and safety of Cesium system was improved (new water storage system, lower pressure, precise water level metering).

\begin{figure}[h!]
\centering
\includegraphics[width=0.4\textwidth]{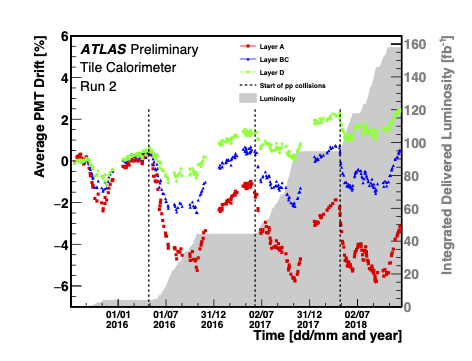}
  \caption{Evolution of the mean relative response of the 3 longitudinal layers (A, BC, D) in the TileCal as a function of time measured by the laser calibration system~\cite{public}.\label{fig:las}}
\end{figure}

\begin{figure}[h!]
\centering
\includegraphics[width=0.4\textwidth]{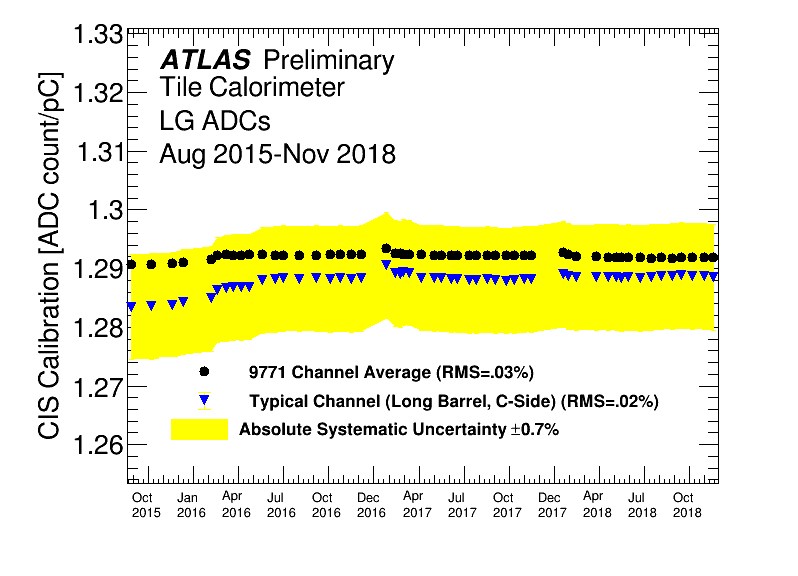}
  \caption{Detector-wide CIS calibration constant averages of all the low-gain ADCs for a selection of CIS calibration runs spanning Run II, plotted as black circles. The CIS constants from a typical channel (LBC20, Channel 33 for LG) are additionally plotted~\cite{public}.\label{fig:cis}}
\end{figure}

\hspace{1pt}
\paragraph{Laser calibration system} \hspace{0pt} \\
The Laser calibration system is used to monitor and measure the gain stability of each PMT by sending a controlled amount of laser light to the photocathode and comparing with a reference light. The response of each channel with respect to its nominal value (at the time of the latest Cesium calibration) is translated into a calibration constant. During the LHC Long Shutdown from 2012 to 2015, a new Laser~II system was developed with new electronics and optical components and light monitoring, which provides an improved resolution. Evolution of the mean relative response as a function of time for different layers during full Run II is shown in Figure~\ref{fig:las}, and the maximal drift is observed in A- and E-cells, which are the cells with highest energy deposits. The precision of the calibration constants is better than 0.5\%.
\hspace{1pt}
\paragraph{Charge injection system} \hspace{0pt} \\
Charge injection system is used to calibrate and monitor the readout electronics. A signal of a known charge is injected to the electronics to evaluate the ADC response over its full charge (approximately 0–800\,pC in low gain and 0–12\,pC in high gain). The conversion factor from ADC counts to pC, is calculated with a linear fit to the peak amplitude versus injected charge and complemented by nonlinear correction factors coming from signal processing. The precision of the calibration system is $\approx 0.7$\% and the variation of the constants are approximately around 0.05\% for individual channels. Detector-wide CIS calibration constants for all low-gain ADCs and for a typical channel (LBC20, Channel 33) during Run II is given in Figure~\ref{fig:cis}.

\hspace{1pt}
\paragraph{Minimum bias integration} \hspace{0pt} \\
LHC proton-proton collisions are dominated by soft parton interactions, which is called \textit{Minimum Bias (MB)} events. The PMT anode current measured by TileCal is proportional to the instantaneous luminosity, and the currents over a time window of about 10~$\mu$s are continuously recorded during the collisions, allowing to monitor and measure the response of the detector to MB events. Therefore, this provides an additional way to measure the luminosity delivered to ATLAS due to the dependency of the currents to the instantaneous luminosity.



\hspace{1pt}
\paragraph{Time calibration and noise} \hspace{0pt} \\
The TileCal status and data quality monitoring includes filtering data corruption and maintaining the per channel timing calibration. Time calibration is performed using the jets and monitored during physics data and the time resolution is found to be better than 1~ns for $E_{cells} > 4~GeV$. There are two main sources for the cell noise in the calorimeter: electronic noise, and pile-up noise, which comes from multiple interactions at the same collision runs. The electronic noise is below 20~MeV for most of the calorimeter cells in Run II. The total noise is increasing with pile-up. The largest noise is in the region with the highest exposure (A- and E -cells).

\begin{figure}[h!]
\centering
\includegraphics[width=0.4\textwidth]{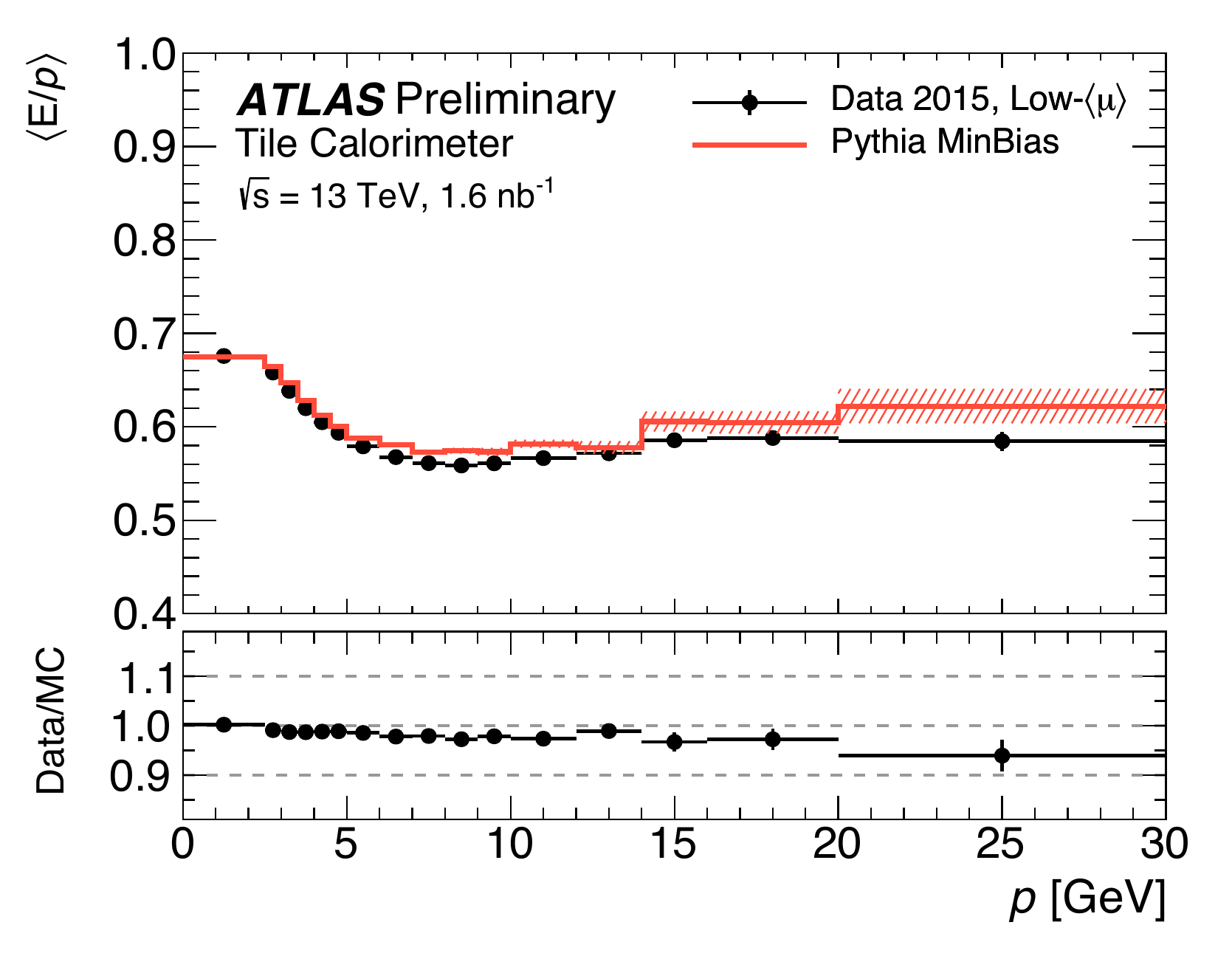}
  \caption{The ratio of the calorimeter energy at EM scale to the track momentum ⟨E/p⟩ of single hadrons in 2015~\cite{public}.\label{fig:sis}}
\end{figure}

\begin{figure}[h!]
\centering
\includegraphics[width=0.4\textwidth]{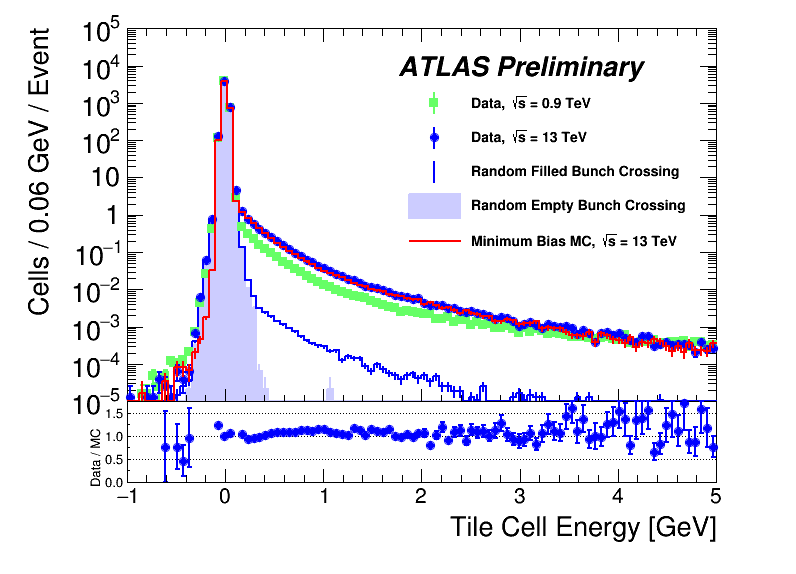}
  \caption{Deposition of energy in TileCal cells for 0.9 and 13 TeV 2015 collision data, Minimum Bias MC, and randomly triggered events~\cite{public}.\label{fig:dep}}
\end{figure}

\hspace{1pt}
    \paragraph{Response to isolated particles and jets} \hspace{0pt} \\
The ratio of the calorimeter energy at EM scale to the track momentum $<E/p>$ of single hadrons is used to evaluate uniformity and linearity during data taking. This ratio is expected to be $<1$ due to the non-compensating nature of the TileCal ($e/h=1.36$) and therefore jets are further calibrated to the jet energy scale. The minimum bias events were used to measure the value of $<E/p>$ and this value agrees with simulation within 5\% (Figure~\ref{fig:sis}). The deposition of total energy in TileCal cells for 0.9 and 13~TeV with 2015 collision data, MinBias Monte Carlo, and randomly triggered events is also shown in Figure~\ref{fig:dep}. A good agreement between the measured contributions of the total cell energy and simulation is found.



\section{Upgrade of the ATLAS Hadronic Tile Calorimeter for the High Luminosity LHC}
The High Luminosity upgrade of the LHC (HL-LHC)~\cite{Apollinari:2116337} will provide an instantaneous luminosity around 7.5 times larger than LHC, which will allow to reach 4000~$fb^{-1}$ dataset. This opens up a new precision physics era in the various physics subjects and provides a probe for Beyond the Standard Model physics with high sensitivity. The HL-LHC is planned to start to provide collisions in 2028, and the ATLAS detector and TileCal must be upgraded to maintain their high performance in challenging HL-LHC environment~\cite{CERN-LHCC-2017-019}. Therefore several TileCal components will be replaced during this period. 

\hspace{1pt}

\paragraph{Mechanics}\hspace{0pt} \\
In order to provide better accessibility, robustness and to reduces the scale of a possible single point of failure, TileCal will change the currently used drawer structures with the structures half as long as the current drawers in Phase II. The new 70 cm-long mechanical structures are called mini-drawers (MD). For the upgraded detector, a super-drawer will consist of 4 independent mini-drawers for LB, each having independent power and signal paths to the back-end electronics. Each MD will host 12 PMTs, 12 Front-End-Cards, 1 Main Board, 1 Daughter Board and 1 HV
passive distribution board. The EB will be consist of 3 mini-drawers and 2 micro-drawers, which are additional drawers that holds the PMTs in the right position and do not include digital electronics. Prototypes of the mini-drawers have been built and validated in a test beam program, and the production of the mechanics is started.

\begin{figure}[h!]
\centering
\includegraphics[width=0.5\textwidth]{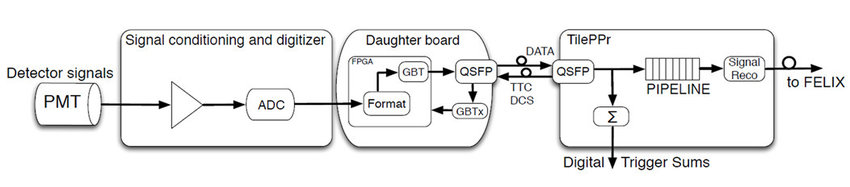}
  \caption{Block diagram of the TileCal readout architecture at the HL-LHC~\cite{CERN-LHCC-2017-019}.\label{fig:HLLHC}}
\end{figure}

\hspace{1pt}
\paragraph{Readout electronics}\hspace{0pt} \\
All TileCal read-out electronics will
be replaced for HL-LHC. The TileCal read-out electronics (Figure~\ref{fig:HLLHC}) is divided into on-detector electronics, where the radiation hardness of the materials must be prioritised, and off-detector electronics located in the counting rooms about 100~m away from the detector.
First of all, around 10\% of the PMTs will be replaced due to ageing, which are located inside the PMT blocks. TileCal introduced a Front End board for the New Infrastructure with Calibration and signal Shaping (FENICS) card, which shapes and amplifies PMT analogue signals. FENICS provides two types of read out: The “fast readout” for physics operates in two different gains in order to cover a dynamic range from 200 fC (MiPs) up to 1000 pC, and the “slow integrator readout”, which integrates the PMT current for calibration of the calorimeter with a Cesium source and for the luminosity measurements. FENICS will provide improved precision and better noise performance. The prototypes are fully qualified for radiation levels expected at HL-LHC, and in the pre-production phase.

A Mainboard (MB) per MD receives data from FENICS from up to 12 PMTs, digitises it, and passes it to the daughterboard. The fast read out uses 24 12-bit ADCs @40Msps, while the integrator readout uses 12 16-bit SAR ADCs. The Mainboard has been fully qualified for the expected radiation environment, and has entered pre-production phase.

A Daughterboard (DB), mounted onto the MB, serves as a link between the on- and off-detector systems. It reads, formats and transfers the data from MB channels to the off-detector systems, while receives and propagates LHC synchronized timing, configuration and control commands to the front-end. Each DB uses 2 Kintex Ultrascale FPGAs. The final prototype is being validated, and all parts for assembly of pre-production is ready, the following batch should include the burn-in process. For robustness, both the MB and the DB have been designed to have two electrically independent sides.

On the off-detector electronics, The PreProcessor (PPr) located in the counting rooms, receives data from the detector, computes energy and time for each PMT channel by using digital filters. Different from the current system, which has on- detector data pipelines, the pipeline buffers are moved to PPr. PPr constitute of the CPMs (Compact Processing Modules) hosted by an ATCA (Advanced Telecommunications Computing Architecture) carrier board, processing the data from two TileCal modules. An ATCA carrier is also arranged to provide a connection between the CPMs and Trigger and Data AcQuisition interface board (TDAQi). The carrier board and CPM have been prototyped in their full size configuration and are being used successfully with a full size demonstrator in ATLAS detector.

\hspace{1pt}
\paragraph{Low and high voltage power supplies}\hspace{0pt} \\
The Tile Calorimeter LVPS (Low Voltage Power Supply) system supplies power to all the front-end electronics of the TileCal. Contrary current 2 stage LV distribution system, the upgraded LV distribution system is composed of 3 stages. Due to their position in the detector, the LVPS are the TileCal component that most exposed to radiation and there have been several radiation test campaigns in order to choose the optimal material. All LVPS bricks convert the input 200~V provided by a circuit supply residing in the USA15 cavern, to provide 10~V for the front-end electronics, which is then converted to different voltages by POL regulators needed by the electronics. It is designed with better reliability, lower noise and improved radiation tolerance and monitoring of the bricks is being done by ELMB (Embedded Local Monitoring Board), hosted by an ELMB motherboard inside the LV box. The ELMB motherboard has been redesigned to be compatible with the new control method of the LVPS bricks. LV system was fully prototyped and tested in vertical slice test with both on and off-detector elements.

The high voltage (HV) distribution system supplies regulated high voltage for each PMTs, which must be stable within better than 0.5~V in the range 600–900~V in order to ensure a 1\% precision on the energy scale in TileCal. High PMT current requires active dividers to keep the linearity even at high luminosity, therefore current passive dividers will be replaced with active dividers. Unlike the current system, HV distribution system will be regulated remotely at the HV regulation board, now called HVremote, is located in the USA15 cavern outside of the radiation environment of the detector. Full size prototypes of the regulation boards have been validated and were shown to meet the HV stability requirements. Full size prototypes of the supply boards are currently being tested with the regulation boards.

\begin{figure}[h!]
\centering
\includegraphics[width=0.4\textwidth]{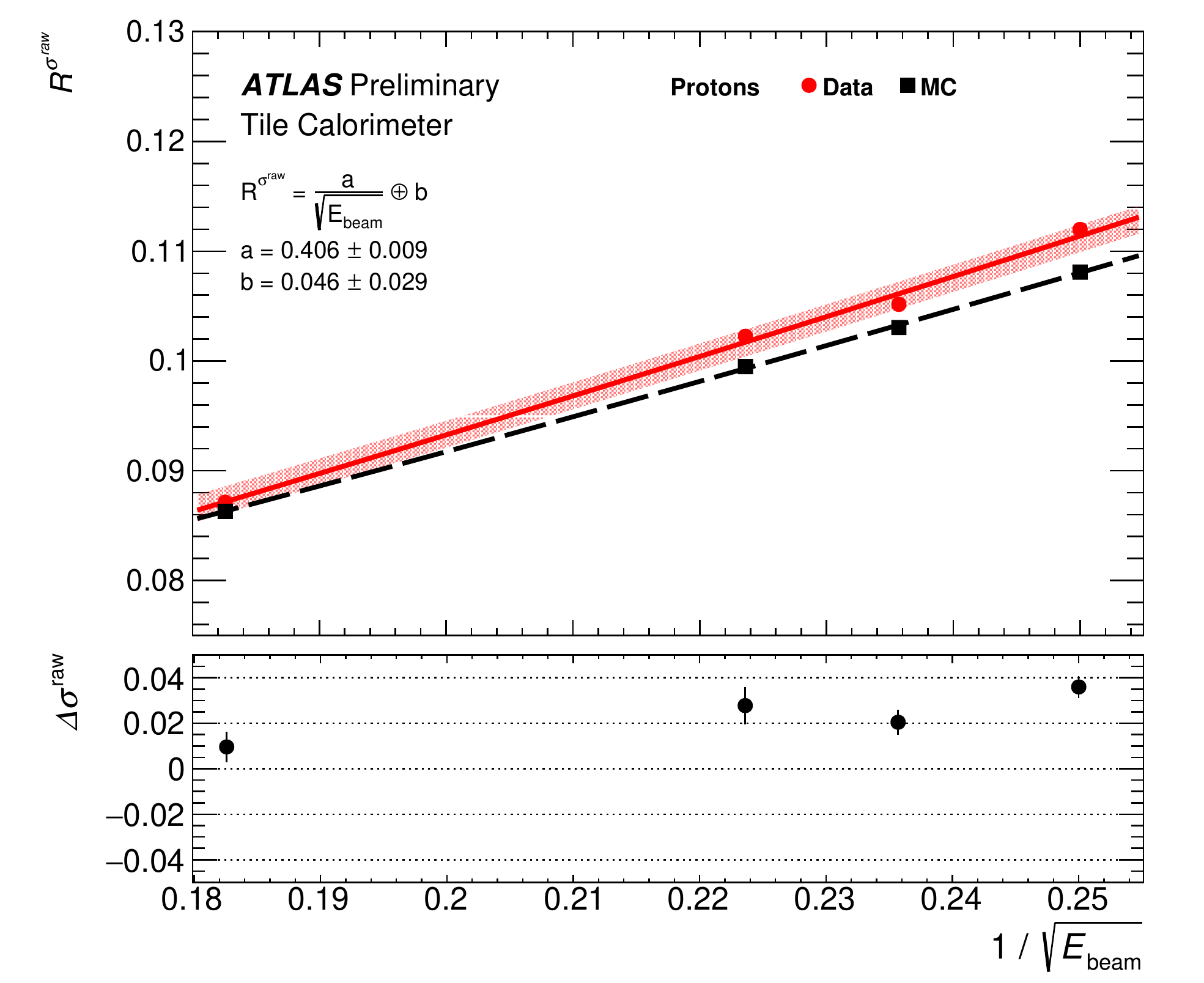}
  \caption{ Fractional resolution, measured (red circles) and predicted by Monte Carlo simulation (black quares) as a function of beam energy obtained in the proton beams~\cite{public}.\label{fig:rsig}}
\end{figure}

\begin{figure}[h!]
\centering
\includegraphics[width=0.4\textwidth]{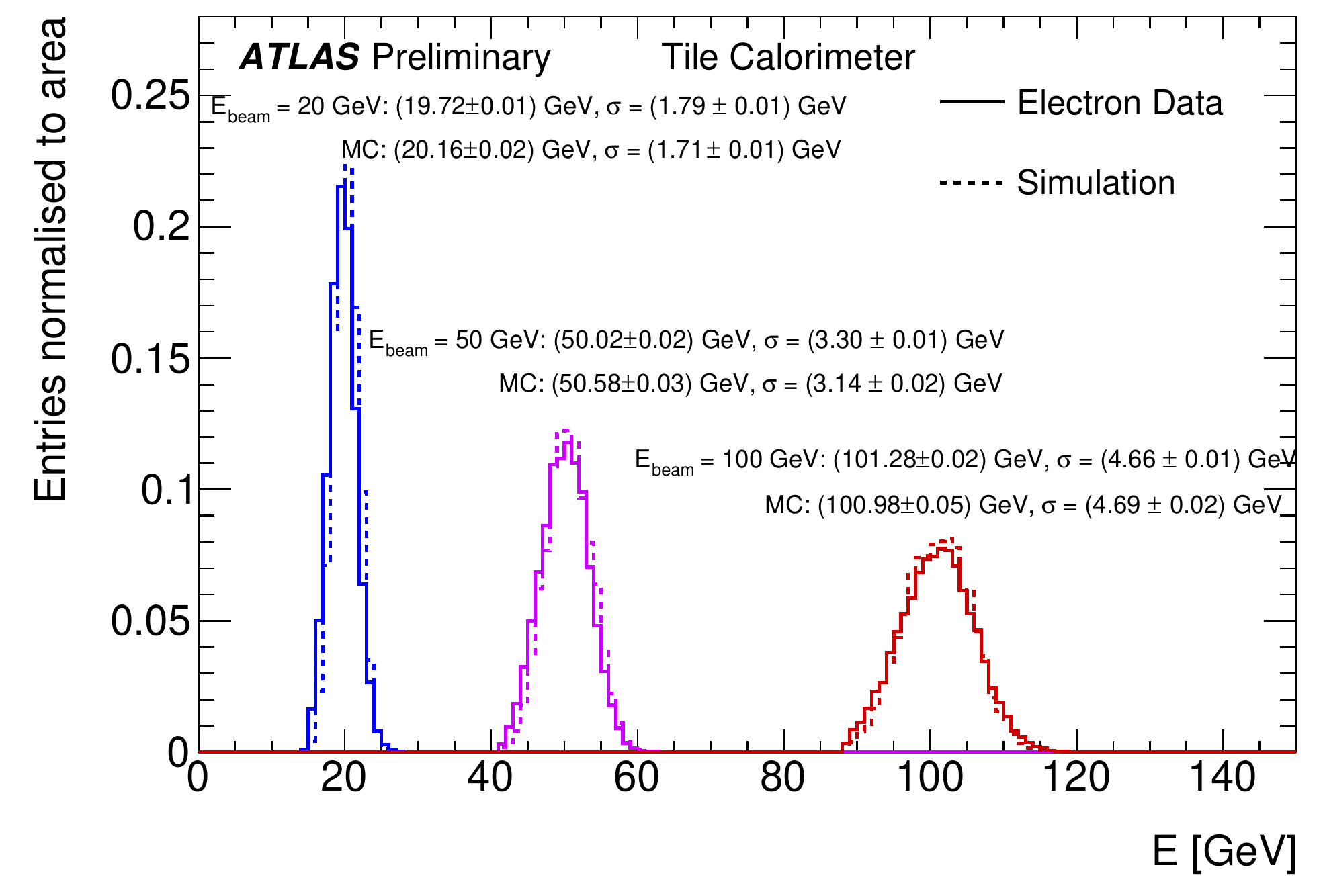}
  \caption{ Distributions of the total energy deposited in the calorimeter obtained using electrons beams of 20, 50 and 100~GeV incident~\cite{public}.\label{fig:ener}}
\end{figure}

\hspace{1pt}
\paragraph{Test beam results}\hspace{0pt} \\
Modules equipped with Phase II upgrade electronics together with modules equipped with the legacy system were exposed to different particles at different energies (muons, electrons, hadrons) in 7 test-beam campaigns at SPS during 2015-2018. Figure~\ref{fig:rsig} shows the fractional resolution ($R^{\sigma_{raw}} = \sigma{raw} / E_{beam}$) of TileCal to proton beams both data and MC as a function of beam energy, and shows the observations agree within the uncertainties. Figure~\ref{fig:ener} shows the distributions of the total energy deposited in the calorimeter obtained using electrons beams of 20, 50 and 100~GeV incident. For a given beam energy the experimental and the simulated shapes are very similar proving the purity of the selected experimental electron samples.


\hspace{1pt}

\newpage

\section{Conclusion}
The Tile Calorimeter is an important part of ATLAS detector at LHC. Sets of calibration systems is used to calibrate and monitor the calorimeter response. Inter calibration and uniformity were monitored with isolated charged hadrons and cosmic muons. The stability of the absolute energy scale at the cell level was maintained to be better than 1\% during Run II.
Wide R\&D program to redesign the on-detector and off-detector electronics for HL-LHC has been done. On- and off-detector electronics will be replaced during Long shutdown in 2025-2027. Readout resolution and sensitivity will be improved slightly. Special attention is given to redundancy, reliability and radiation hardness. Many components of the mechanics and front-end electronics have entered pre-production or production.


%

\ifCLASSOPTIONcaptionsoff
  \newpage
\fi



%

\end{document}